\documentclass[10pt]{article} 
\begin{document} 
\baselineskip=10pt \def\no{\noindent} 
\def\be{\begin{eqnarray}} \def\ee{\end{eqnarray}} \def\non{\nonumber} 
\title{BFFT Formalism Applied to the Minimal Chiral Schwinger Model} 
\author{C. P. Natividade,$^{1,2}$ 
H. Boschi-Filho$^3$\thanks{e-mail: boschi@if.ufrj.br} and 
L. V. Belvedere$^{1}$\thanks{e-mail: belve@if.uff.br}\\ 
\it\small $^1$Instituto de F\'\i sica, Universidade Federal Fluminense\\ 
\it\small Avenida Litor\^anea 
s/n, Boa Viagem, Niter\'oi, 24210-340 Rio de Janeiro, Brazil\\ 
\it\small $^2$Departamento de F\'\i sica e Qu\'\i mica, 
Universidade Estadual Paulista\\ 
\it\small Avenida Ariberto Pereira da Cunha 333, Guaratinguet\'a, 
12500-000 S\~ao Paulo, Brazil\\ 
\it\small $^3$Instituto de F\'\i sica, 
Universidade Federal do Rio de Janeiro\\ 
\it\small Caixa Postal 68528, Rio de Janeiro, 
21945-970 Rio de Janeiro, Brazil} 
\date{\today} 
\maketitle 
\begin{abstract} 
We consider the minimal chiral Schwinger model, by embedding the gauge 
noninvariant formulation into a gauge theory following the 
Batalin-Fradkin-Fradkina-Tyutin point of view. Within the BFFT procedure, 
the second class constraints are converted into strongly involutive 
first-class ones, leading to an extended gauge invariant formulation. 
We also show that, like the standard chiral model, in the minimal chiral 
model the Wess-Zumino action can be obtained by performing a q-number 
gauge transformation into the effective gauge noninvariant action. 
\end{abstract} 


\section{Introduction} 
\renewcommand{\theequation}{1.\arabic{equation}} 
\setcounter{equation}{0} 
Following Dirac's conjecture a critical issue in the study of a gauge 
model is the presence of first-class constraints \cite{Dirac}. 
First-class constraints are related to symmetries while the second-class 
ones may imply some ambiguities when treated as quantum field operators. 
The physical status of a theory is chosen by imposing complementary 
conditions which are given by the first-class constraints. 
The presence of second-class constraints is in general avoided. 
There are many procedures which allow the exclusion of these constraints 
from the effective action \cite{Henneaux}. One of them is the 
Batalin-Fradkin-Fradkina-Tyutin (BFFT) method \cite{Batalin,BFT89,BT91}, 
which converts second-class constraints into first-class ones by 
introducing auxiliary fields. 
The BFFT formalism has been employed in different models, as for 
example, the chiral boson model \cite{Amorim}, the massive Maxwell and 
Yang-Mills theories \cite{Fujikawa,Banerjee97,KimRothe}, the CP$^{N-1}$ 
model \cite{Banerjee94}, the non-linear sigma model \cite{barcelos97}, 
the chiral Schwinger model \cite{Kim97,Park98} and more recently a fluid 
field theory \cite{NB00}. As expected, the implementation of the BFFT 
method through the introduction of new fields gives rise to a kind of a 
Wess-Zumino term which turns the resulting extended theory gauge invariant. 
In particular, an elegant way of obtaining the Wess-Zumino term and the 
effective action is the BRST-BFV procedure \cite{Henneaux85}. 

On the other side, two dimensional models have played an important role 
in theoretical physics as a laboratory where many interesting phenomena 
can be studied in a fashion which is usually easier to handle than more 
realistic four dimensional theories \cite{AAR}. One well known model is 
the two dimensional quantum electrodynamics ($QED_2$) which was 
introduced long ago by Schwinger \cite{Schwinger} to discuss dynamical 
mass generation for gauge fields without breaking the gauge symmetry 
\cite{Swieca}. 
More recently, Jackiw and Rajaraman proposed a model \cite{Rajaraman} 
with a chiral coupling between the two dimensional gauge and fermion 
fields mimicking the weak interactions of the standard model. 
This two dimensional model, known as the chiral Schwinger model, happens 
to be gauge anomalous although unitary and carries an arbitrary 
regularization parameter $a$ in all of its physical 
quantities (mass, propagator, etc) (see, {\sl e. g.,} 
\cite{Carvalhaes,Bel} and references therein). 
Another interesting two dimensional chiral model is the one that describes 
right or left movers, {\sl i. e.}, chiral bosons which were introduced by 
Siegel \cite{Siegel} inspired on the heterotic string and later reobtained 
by Floreanini and Jackiw \cite{Floreanini}. 
In particular, Harada \cite{Harada,Harada2} considered a version of the 
chiral Schwinger model but without the right-handed fermions. 
He showed that this model known as the minimal chiral Schwinger model 
corresponds to a gauged version of the Floreanini-Jackiw chiral boson. 
Naturally, this model share some properties with the complete chiral model 
as the gauge anomaly and the dependence on an arbritrary parameter 
(which is usually called $a$, as in the original Jackiw-Rajaraman model) 
and the novelty here is the description of chiral bosons which in some 
sense represents the motion of superstrings. 
Furthermore, a left-handed Wess-Zumino (WZ) action has been built for this 
model by considering an antichiral constraint \cite{Kye91,Kye,Dutra}. 

In this paper we discuss the Hamiltonian formalism for the minimal chiral 
Schwinger model using the BFFT method. The application of this method 
introduces in a very natural way the chiral constraints in the model. 
Otherwise, one would be forced to put these constraints by hand. 
Then, using the BRST-BFV procedure we obtain the Wess-Zumino term with a 
set of the first-class constraints and the total effective action. 

We also show that, like the standard chiral model, in the minimal chiral 
model the WZ action can be obtained by performing a q-number gauge 
transformation into the effective gauge non-invariant (GNI) action. 

This paper is organized as follows: In section II we discuss the 
conversion of the second-class to first-class constraints, by using the 
BFFT method for the miminal chiral Schwinger model. 
We obtain the corresponding extended Hamiltonian in strong 
involution with the first class constraints. 
In section III, we obtain the extended gauge invariant effective 
action which brings in a Wess-Zumino term. 
In section IV we discuss the generation of the WZ action by 
performing a q-number gauge transformation on the gauge non-invariant 
effective action. 
Finally, in section V we make some remarks about the ``fermionization'' 
of the extended gauge invariant formulation of the anomalous model for 
$a = 2$ and give general arguments contrary to the equivalence to the 
vector Schwinger model advocated in the literature \cite{Kye,Carena}. 
An appendix is also included where we present some details on 
the calculation of the extended canonical Hamiltonian. 


\section{Extended First-Class Hamiltonian} 
\renewcommand{\theequation}{2.\arabic{equation}} 
\setcounter{equation}{0} 

In order to implement the canonical BFFT scheme, it is necessary to 
specify the Hamiltonian together with the set of the constraints of the 
model \cite{Batalin,BFT89,BT91}. Here we are going to apply the general 
BFFT method following the lines of work in Refs. 
\cite{Fujikawa,Banerjee97,KimRothe}. 

To begin with, let us start by considering the bosonized version of the 
minimal chiral Schwinger model \cite{Harada,Harada2}, described by the 
following Lagrangian density \cite{Kye91,Kye,Dutra} 
\be {\cal L} [ \phi, A_\mu ] 
= {\cal L}_M [ A_\mu ] + \check {\cal L} [ \phi, A_\mu ]\,, 
\ee 
\no where the Maxwell Lagrangian is, 
\be 
{\cal L}_M = - \frac 14 F_{\mu\nu}F^{\mu\nu}\,, 
\ee 
\no the gauge noninvariant (GNI) contribution is given by, 
\be
\label{l1} 
\check {\cal L} [ \phi, A_\mu ] 
&=& \, \dot\phi\, \, \phi^\prime
 - (\, \phi^\prime)^2 + 2\, e \, \phi^\prime (A_0-A_1) 
- \frac 12\, e^2 \,  (A_0-A_1)^2 \non\\ & & + \frac 12\, a \,\, e^2 \,  
\Big ( (A_0)^2 - (A_1)^2 \Big ) \,. 
\ee 
\no with the Jackiw-Rajaraman parameter $a > 1$, 
overdot means partial time derivative 
($\, \dot\phi=\partial_0\phi=\partial\phi/\partial t 
= \partial^0\phi$) and prime denotes partial space derivative 
($\, \phi^\prime=\partial_1\phi=\partial\phi/\partial x^1 
=-\partial^1\phi$). 
The canonical momenta are given by 
\be &&\Pi^\mu = \frac{\partial{\cal L}}{\partial 
\dot{A}_\mu} = F^{\mu0}; \\ 
&&\Pi_\phi = \frac{\partial{\cal L}}{\partial 
\dot{\phi}} = \, \phi^\prime \,, 
\ee 
\no which imply the primary constraints 
\be 
&&\Omega_1 = \Pi^0 \approx 0\,; \label{Omega1} \\ 
&& \Omega_2 = \Pi_\phi 
-\, \phi^\prime \approx 0\,. \label{Omega2} 
\ee 
\no The corresponding canonical Hamiltonian is given by, 
\be H_c &=& \int {\rm d}x^1 \Big\{ \frac 12 (\Pi^1)^2 + 
(\, \phi^\prime)^2 - 2\, e \, \phi^\prime (A_0-A_1) \non\\ && \qquad 
+ \frac 12\, e^2 \,  (A_0-A_1)^2 - \frac 12\, a \,\, e^2 \,  (A_0)^2 \non\\ 
&& \qquad + \frac 12\, a \,\, e^2 \,  (A_1)^2 + 
\Pi^1 \partial_1 A_0 \Big\}\,, \label{Hc} 
\ee 
\no where we are using the conventions 
$F^{10}=\Pi^1= E^1=\partial^1A^0 -\partial^0A^1=\dot{A_1}-\partial_1 A_0$. 
Time conserving of the primary constraints lead to the Gauss law 
\be 
\Omega_3 &=& \dot{\Omega}_1=\left\{\Pi^0,H_c\right\} \non\\ 
&=& \partial_1 \Pi^1 -\, e \, J^0 \approx 0\,, \label{Omega3} 
\ee 
\no where the current is given by, 
\be J^0 = 2 \, \phi^\prime +\, e \, [(a-1) A_0 + A_1]\,. 
\ee 
\no The system given by the Poisson brackets $\{\Omega_1,\Omega_3\}$ 
constraints is second-class. The time evolution of $\Omega_3$ does not 
lead to any new constraints but determines the Lagrange multiplier of the 
$\Omega_1$ constraint. So, the algebra of the constraints is given by a 
set 
$\{\Omega_j\}$ which can be determined using the BFFT scheme. 
In order to simplify this procedure we shall implement the constraints 
$\Omega_j=0$ strongly by introducing Dirac brackets 
\cite{Banerjee97,KimRothe}. Through the Dirac's procedure we have that 
\be 
\left\{\Omega_i,\Omega_j\right\}^D=0\,, 
\ee 
\no and the remaining 
\be 
\left\{\chi_i(x),\chi_j(y)\right\}^D=\,\Delta_{ij}(x,y)\,, 
\ee 
\no where we defined $\chi_1=\Omega_1$ and $\chi_2=\Omega_3$, from now on 
$x\equiv x^1$, $y\equiv y^1$ and 
\be 
\label{Delta} 
\Delta_{ij}(x,y) = \left(\begin{array}{cc}0 
&\, e^2 \, (a-1)\\ -e^2(a-1) & 2e^2\partial_{x} \end{array}\right) 
\delta(x-y)\,. 
\ee 

In order to reduce the second-class system to a first-class one, 
we begin by extending the phase space including the new fields 
$\theta_i(x)$ which satisfy the algebra: 
\be 
\left\{\theta_i(x),\theta_j(y)\right\}^D=-\, \epsilon_{ij} \, 
\delta(x-y)\,, 
\ee 
\no where $\epsilon^{12}=-\, \epsilon_{12}\, =+1$. 

The first-class $\tilde\chi_i$ are now constructed as power series 
\cite{Fujikawa,Banerjee97,KimRothe} 
\be 
\label{tildechi} 
\tilde\chi_i = \chi_i + \sum_{n=1}^\infty \chi_i^{(n)}, 
\ee 
\no where $\chi_i^{(n)}$ are homogeneous polynomials of order $n$ in 
the auxiliary fields $\theta_i(x)$, to be determined by the requirement 
that the constraints $\tilde\chi_i$ be strongly involutive 
\be 
\left\{\tilde\chi_i(x),\tilde\chi_j(y)\right\}^D=0. 
\ee 

The first-order correction for the expression (\ref{tildechi}) can be 
written as 
\be 
\label{tildechi2} 
\tilde\chi_i = \chi_i + \int {\rm d}y\ \sigma_{ij}(x,y)\ 
\theta_j(y), 
\ee 
\no where the quantities $\sigma_{ij}(x,y)$ are implicitly defined by 
\be 
\Delta_{ij}(x,y)=\int {\rm d}z\ {\rm d}z^\prime \sigma_{ik}(x,z) 
\, \epsilon_{kl}\, \sigma_{jl}(z^\prime,y), 
\ee 
\no with $\Delta_{ij}(x,y)$ given by eq. (\ref{Delta}). 
By performing the calculations and choosing 
$\sigma_{ij}(x,y)$ such that $\tilde\chi_i$ are linear in the fields 
$\theta_i(x)$, we obtain, 
\be 
\label{sigma} \sigma_{ij}(x,y) = 
\left(\begin{array}{cc} 1 & 0 \\ \frac1{a-1}\partial_{x} & 
-e^2(a-1)\end{array}\right) \delta(x-y). 
\ee 
\no Consequently, we get, 
\be 
&&\tilde\chi_1=\chi_1+\theta_1(x) \non\\&&\tilde\chi_2=\chi_2+\frac 
1{a-1}\partial_{1}\theta_1 -\, e^2 \, (a-1) \theta_2, 
\ee 
\no which are first-class. 
The above set permit us to compute the extended first-class Hamiltonian, 
\be 
\tilde H = \sum_{n=0}^\infty H^{(n)}, 
\ee 
\no where $H^{(n)}\sim\theta_n$, with the subsidiary condition 
\be 
H^{(0)} \equiv H_c. 
\ee 

The general expression for the iterated Hamiltonian $H^{(n+1)}$ is 
given as a recurrence relation as in Refs. 
\cite{Fujikawa,Banerjee97,KimRothe}, 
\be 
H^{(n+1)}&=& - \frac {1}{n+1} 
\int {\rm d}x\ {\rm d}y\ {\rm d}z\ \theta_\alpha(x) 
\left(\omega_{\alpha\beta}\right)^{-1} 
\left(\sigma_{\beta\gamma}\right)^{-1} 
G_\gamma^{(n)}, \ee \no where \be \label{Gn} 
G_\gamma^{(n)}=\left\{\chi_\gamma,H^{(n)}\right\}. 
\ee 
\no Here, we mention that $\theta_n=0$, for $n\geq 2$. 
Since $(\omega_{\alpha\beta})^{-1}$ and $(\sigma_{\beta\gamma})^{-1}$ 
are proportional to Dirac delta functions and considering the canonical 
Hamiltonian, eq. (\ref{Hc}), we get, 
\be 
G_1^{(0)}&=&\left\{\chi_1,H^{(0)}\right\}\non\\ &=& \chi_2 
\non\\G_2^{(0)}&=&\left\{\chi_2,H^{(0)}\right\}\non\\&=& 
e^2\left[(a-1)\partial_{1}A_1 - \Pi^1\right]. 
\ee 
\no Performing the shift in the fields 
\be 
\label{shift1} 
\theta_1(x) &\longrightarrow & e(a-1)\, \theta \\ 
\label{shift2}
\theta_2(x) &\longrightarrow & \frac 1{e(a-1)}\Pi_\theta, 
\ee 
\no 
we obtain the first-order correction for the canonical Hamiltonian 
\be 
H^{(1)} 
&=& - \int {\rm d}x \Big\{\frac 1{e(a-1)}\left(\, \theta^\prime + 
\Pi_\theta\right)\chi_2\non\\&&\qquad\quad+\left[(a-1)\partial_1 A_1 - 
\Pi^1\right]\, \theta \Big\}, \label{Hc1} 
\ee 
\no where $\, \theta^\prime\equiv\partial_1\theta$. Following the same 
steps leading to the first-order corrections we obtain the second-order 
Hamiltonian (see the Appendix A) 
\be 
H^{(2)} &&= - \frac 12 \int {\rm d}x \left[ \frac 1{a-1}(\Pi_\theta)^2 - 
\beta (\, \theta^\prime)^2 +\, e^2 \,  \theta^2 \right], 
\label{Hc2} 
\ee 
\no with $\beta=(a-1)+(a-1)^{-1}$. Putting together the results from Eqs. 
(\ref{Hc}), (\ref{Hc1}) and (\ref{Hc2}), we find the extended Hamiltonian 
\be 
\tilde H = H^{(0)} + H^{(1)} + H^{(2)}, 
\ee 
\no which is strongly involutive with respect to the constraints 
$\tilde\chi_i(x)$. On the other hand, an inspection of the complete set 
of constraints reveals that 
\be 
\left\{\tilde\chi_i,\tilde\chi_j\right\}=0, 
\ee 
\no with $i,j=1,2,3$. These results clearly illuminate the first-class 
nature of the system. The next step is to calculate the effective action 
which should be invariant under extended gauge transformations, as we are 
going to show in the following section.


\section{Effective Gauge Invariant Action} 
\renewcommand{\theequation}{3.\arabic{equation}} 
\setcounter{equation}{0} 

Let us obtain the effective action through the BRST-BFV 
formalism \cite{Henneaux85}. This method permit us to obtain the effective 
action in a direct way by including Lagrange multipliers and ghost fields 
with the corresponding canonical momenta and a gauge fixation function 
which together with the BRST charge operator generate the terms that lead 
to the expected gauge invariant action. 

Therefore, following the usual BFV prescription and considering the 
Eqs. (\ref{Hc1}), (\ref{Hc2}), the effective action can be written as, 
\be 
S_{eff}&=&\int {\rm d}^2 x \Big\{\Pi^0\, \dot A_0 +\Pi^1\, \dot A_1 + 
\Pi_\phi\, \dot\phi + \Pi_\theta \, \dot\theta - {\cal H}^{(0)} 
\non\\ 
&& +\frac 1{e(a-1)}(\Pi_\theta+\, \theta^\prime)\chi_2 
-e\left[(a-1)\partial_1A_1 + \Pi^1\right]\, \theta \non\\ && -\frac 
1{2(a-1)}(\Pi_\theta)^2 
+ \frac 12\beta(\, \theta^\prime)^2 
-\frac 12e^2\theta^2\Big\}
\non\\ 
&&+\int {\rm d}^2x \left[\dot\lambda_a p_a + 
\bar{\cal P}_a \, \dot c_a +{\dot{\bar c}}_a {\cal P}_a 
+\left\{\Psi,Q\right\}\right], 
\label{Seff} 
\ee 
\noindent where $(c_a,\bar{{\cal P}_a})$ and $({\cal P}_a,\bar{c_a})$ 
form a pair of canonical ghost-antighost fields with opposite Grassmanian 
parity 
\be 
\{c_a(x),\bar{{\cal P}_b}(y)\}=\{{\cal P}_a(x),\bar{c_b}(y)\} 
= \,\delta_{ab}\,\delta(x-y), 
\ee 
\no while $(\lambda_a,p_a)$ is a canonical Lagrange multiplier set 
\be 
\{\lambda_a(x),p_b(y)\}=\,\delta_{ab}\,\delta(x-y). 
\ee 
\no The charge operator $Q$ is defined as 
\be 
Q=c_a\tilde\chi_a +p_a{\cal P}_a, 
\ee 
\no with $\tilde\chi_a$ being the first-class constraints, as discussed 
in the previous section. Finally, the fermion operator $\Psi$ is 
\be 
\Psi={\bar c}_a\alpha_a + {\bar{\cal P}}_a\lambda_a, 
\ee 
\no where $\alpha_a$ are the Hermitian gauge-fixing functions. 
Different choices of the gauge functions $\alpha_a$ can be done in 
order to obtain the effective action. The partition function is then 
given by 
\be 
{\cal Z} = \int [ {\cal D} \Sigma]\, e^{\,i\,S},\label{Z} 
\ee 
\no where the functional integral measure $[{\cal D} \Sigma]$ 
includes all the fields appearing in the action (\ref{Seff}). 

Before going on, we can make the scaling $\;\alpha_a\to\alpha_a/M\;$, 
$\;p_a\to Mp_a\;$ and $\;\Sigma_a\to M\Sigma_a\;$, in such a 
way that the Jacobian of this transformation is equal to the unity. 
One can then verifies that in the limit $\;M \to 0\;$, the action is 
independent of ghost and antighost fields. 

Now, we can perform the choice of the gauge function and do 
some of the integrations implied in $[{\cal D} \Sigma ]$. 
To this end, we choose 
\be
\label{gc1} 
&&\alpha_1(x)=\Pi_\phi-\, \phi^\prime + \dot{\lambda}_1 \\ 
&&\alpha_2(x)=\Pi_\theta+\, \theta^\prime + e(a-1)A_0+eA_1 
+\dot{\lambda}_2.\label{gc2} 
\ee 

Since the first-class constraints are 
\be 
\tilde{\chi_1}&=&\chi_1+\,e\,(a-1)\, \theta \\ 
\tilde{\chi_2}&=&\chi_2+\,e\,(\, \theta^\prime-\Pi_\theta), 
\ee 
\no where $\chi_1\equiv\Omega_1$ and $\chi_2\equiv\Omega_3$, 
given by Eqs. (\ref{Omega1}), (\ref{Omega2}) and (\ref{Omega3}), 
we can compute the Poisson bracket 
\be 
\{\Psi,Q\} &=& -[\Pi_\phi-\, \phi^\prime + \dot{\lambda}_1]p_1 \non\\ && 
-[\Pi_\theta+\, \theta^\prime + \, e\, (a-1)\, A_0\, 
+\, e\, A_1\, +\dot{\lambda}_2]\, p_2 
\non\\ && 
-[\Pi_0+\,e\,(a-1)\,\theta]\lambda_1 \non\\ && 
-[\chi_2+\,e\,(\, \theta^\prime-\Pi_\theta)]\lambda_2. \label{PsiQ} 
\ee 

The dynamical terms $\dot{\lambda}_ap_a$ which appear in Eq. (\ref{PsiQ}) 
are cancelled by the similar ones in the original action. After 
integrations over the fields $(p_1,p_2)$ and $(\lambda_1,\lambda_2)$, 
we arrive at the delta functionals 
$\delta(\Pi_\theta+\, \theta^\prime + e(a-1)A_0+eA_1)$, 
$\delta(\Pi_\phi-\, \phi^\prime)$, $\delta(\Pi_0+e(a-1)\theta)$ and 
$\delta(\chi_2+e\, \theta^\prime-e\Pi_\theta)$ in the partition function 
(\ref{Z}). Performing the integrations over $(\Pi_\phi,\Pi_\theta)$ and 
$(\Pi^0,\Pi^1)$ and using the fact that the Hamiltonian $H^{(0)}$ is 
quadratic in the field $\Pi^1$, we get the gauge invariant (GI) 
effective action
\be
\label{egia} 
S_{eff}^{GI} [ \phi, \theta, A_\mu ] = \check S [ \phi, 
A_\mu ] + S [\theta, A_\mu ]\,, 
\ee 
\no where $\check S [ \phi, A_\mu ]$ is the action corresponding to the 
Lagrangian (\ref{l1}), 
\be 
\check S [\phi, A_\mu ] 
&=& \int {\rm d}^2x \Big\{ \, \dot\phi\, \phi^\prime 
- (\, \phi^\prime)^2 + 2e\, \phi^\prime(A_0-A_1) -\frac 12\, e^2 \, (A_0-A_1)^2 
\non \\
&& + \frac 12\, e^2 \, a[(A_0)^2-(A_1)^2]\Big \}\,, 
\label{Sphi} 
\ee 
\no and $S [\theta, A_\mu ]$ is given by, 
\be 
\label{Stheta} 
S [\theta, A_\mu ] = S_{_{WZ}} [ \theta, A_\mu ] - \frac 12 
\frac {e^2}{a-1}\,\int d^2 x [(a-1)A_0+A_1]^2\,, 
\ee 
\no with the WZ action given by, 
\be
\label{WZA} 
S_{_{WZ}} [ \theta, A_\mu ] = \int {\rm d}^2x 
\Big\{ -\dot\theta\, \theta^\prime -\frac 12 \beta(\, \theta^\prime)^2 
+ \,e\,\beta A_1\, \theta^\prime\,+\,2\,e\,A_0\, \theta^\prime \Big\}\,. 
\ee 
\no The effective GI action can be rewritten as, 
\be
\label{effgiaction} 
S_{eff}^{GI} [ \phi, \theta, A_\mu ] 
= S_{eff}^{GNI} [ \phi, A_\mu ] \,+\,S_{_{WZ}} [ \theta, A_\mu ]\,, 
\ee 
\no where the GNI action is given by, 
\be
\label{egnia} 
S_{eff}^{GNI} [ \phi, A_\mu ] 
= \,\int {\rm d}^2x \,\Big\{ \, \dot\phi\, \phi^\prime - 
(\, \phi^\prime)^2 \,+\, 
2e\, \phi^\prime(A_0-A_1)\,-\,\frac{1}{2}\,e^2\,(\beta + 2)
\,(A_1)^2 \Big \} 
\ee 
\no The effective action (\ref{effgiaction}) is invariant under extended 
gauge transformations, 
\be
\label{egt1} 
^g\phi = \phi + g\,, \ee \be\label{egt2} 
^g\theta = \, \theta - g\,, \ee \be\label{egt3} ^gA_\mu 
= A_\mu - \frac{1}{e}\,\partial_\mu g\,. 
\ee 
\no This could be achieved with the use of the first-class constraints 
$\tilde\chi_1$, $\tilde\chi_2$, representing the gauge symmetry of the 
model, introduced by the use of the BFFT method as discussed in the 
previous section. The generators of the symmetry transformations can be 
written as \cite{Galvao} 
\be 
G = \int {\rm d}x \left( a_1\tilde\chi_1 + a_2\tilde\chi_2 \right), 
\ee 
\no where the coefficients $a_j$ are determined through the relations 
$\;\{\chi_1,H_c\}=a_1\chi_1\;$ and $\;\{\chi_2,H_c\}=a_2\chi_2\;$. 
From Eq. (\ref{Omega3}), we obtain 
\be 
\delta A_1 &=& \{A_1,G\}\epsilon \non\\ 
&=& - \frac 1e \partial_1\epsilon\,, 
\ee 
\no and similarly $\delta\phi=\epsilon=-\delta\theta$, where 
$\epsilon= \epsilon(x)$ is a gauge parameter.


\section{Operator Gauge Transformation and the WZ Action} 


As was stressed in Ref. \cite{BR} for the standard chiral $QED_2$ and 
in \cite{BRR} for the chiral $QCD_2$, the WZ action can be obtained via 
an operator-valued gauge transformation of its GNI effective quantum 
action. The resulting extended GI theory is isomorphic to original GNI 
theory. The isomorphism between these two formulations is valid in 
an arbitrary gauge. In what follows we show that in the minimal chiral 
model the GI formulation can also be obtained by performing a gauge 
transformation on the GNI formulation. To begin with, let us consider 
the q-number gauge transformation 
\be
\label{gt1} 
\phi\,\rightarrow\, ^{\theta}\!\phi = \phi + 
\theta\,, \ee \be\label{gt2} A_\mu\,\rightarrow\,^{\theta}\!A_\mu 
= A_\mu - \frac{1}{e}\,\partial_\mu \theta\,, 
\ee 
\no acting on the effective GNI action given by (\ref{egnia}), 
\be
 ^\theta S_{eff}^{GNI} [ \phi, A_\mu ] 
&=& \int {\rm d}^2x \,\Big\{ ( \, \dot\phi + \, \dot \theta ) 
( \, \phi^\prime + \, \theta^\prime ) 
- (\, \phi^\prime + \, \theta^\prime )^2 
\non \\
&&+ 2e ( \, \phi^\prime + \, \theta^\prime ) 
\Big( A_0-A_1 - \frac{1}{e} (\, \dot \theta - \, \theta^\prime )\Big) 
\non \\
&& -\; \frac{1}{2}\,e^2 ( \beta + 2 )
\;(A_1 - \frac{1}{e}\,\theta^\prime )^2\,. 
 \label{gta} 
\ee 
\no The gauge transformed action (\ref{gta}) is manifest invariant under 
extended gauge transformations (\ref{egt1}),(\ref{egt2}), (\ref{egt3}), 
and is identical to the extended GI effective action given by 
(\ref{egia}), 
\be 
^\theta S_{eff}^{GNI}[ \phi, A_\mu ] 
= S_{eff}^{GI} [ \phi, A_\mu, \, \theta ] 
= S_{eff}^{GNI} [ \phi, A_\mu ]\,+\, S_{WZ}[ \theta, A_\mu ]\,. 
\ee 

The WZ action (\ref{WZA}) can also be obtained by performing a q-number 
gauge transformation on the bosonized action corresponding to the 
original GNI Lagrangian (\ref{l1}), 
\be 
\check S [ \phi, A_\mu ] 
&=& \int d^2 x \Big \{\,\dot\phi\, \phi^\prime 
- (\, \phi^\prime)^2 + 2\, e \, \phi^\prime (A_0-A_1) 
\non \\
&& -  \frac 12\, e^2 \,  (A_0-A_1)^2 + 
\frac 12\, a \,\, e^2 \,  \Big ((A_0)^2 - (A_1)^2 \Big )\Big \} \,. 
\ee 

\no In this way, using (\ref{gt1})-(\ref{gt2}), we obtain 
\be
\label{LGI} 
^\theta \check S [\phi, A_\mu ] 
= \check S [\phi, \theta, A_\mu ] = \check S [\phi, A_\mu ] + \check 
S_{WZ} [\theta, A_\mu ]\,, 
\ee 
\no where the WZ action is now given by, 
\be  
\check S_{WZ}[\theta, A_\mu ] &=& 
\int d^2 x \Big \{ \frac{1}{2} (a - 1)\, (\, \dot \theta )^2 
- \frac{1}{2} (a - 1)( \, \theta^\prime )^2 
\non\\
&& +\, e \, A_0\,\Big ( \, \theta^\prime - (a - 1)\, \, \dot \theta \Big ) 
+\, e \, A_1\,\Big ( (a - 1)\, \theta^\prime - \,\dot \theta \Big )\Big \}\,. 
\label{wza2}
\ee 
\no The action (\ref{wza2}) is the usual WZ action obtained for the 
standard chiral model. In this case, the canonical momentum associated 
with the field $\theta$ is given by, 
\be
\label{cm} 
\Pi_\theta = (\, a \, - 1 )\, \, \dot \theta -\, e \, [ (a - 1) A_0 + A_1 ]\,. 
\ee 
\no The effective GI action obtained in the previous section using the 
BRST-BFV formalism, is gauge fixed by the gauge conditions 
(\ref{gc1})-(\ref{gc2}). In order to map the WZ action (\ref{wza2}) 
into (\ref{WZA}), we must impose on (\ref{wza2}) the following condition 
\be
\label{gc} 
\, \theta^\prime + (a - 1) \, \, \dot \theta \approx 0\,. 
\ee 
\no The condition (\ref{gc}) play the role of the delta function 
$\delta ( \Pi_\theta + \, \theta^\prime +\, e \, (a - 1)A_0 + A_1 )$ that 
appears in the integrations over the fields $(p_1,p_2)$ and 
$(\lambda_1, \lambda_2)$ performed in the previous section to obtain 
the effective action. 
Under the condition (\ref{gc}) the canonical momentum (\ref{cm}) is 
mapped into 
\be
\label{cm1} 
\Pi_\theta = - \, \theta^\prime -\, e \, [ (a - 1) A_0 + A_1 ]\,. 
\ee 
\no Indeed, rewriting the standard WZ action (\ref{wza2}) as, 
\be 
\check S_{WZ}[\theta, A_\mu ] &=& 
\int d^2 x \,\Big \{ (a - 1) (\, \, \dot \theta )^2 - 
\frac{1}{2} (a - 1) [ (\, \, \dot \theta )^2 + ( \, \theta^\prime )^2 ]
 \Big \} 
\non \\
&& +\, e \, A_0\,\Big ( \, \theta^\prime - (a - 1) \, \dot \theta \Big ) 
+ e\; A_1\,\Big ( (a - 1) \, \theta^\prime - \, \dot \theta \Big )\Big \}\,, 
\ee 
\no and using (\ref{gc}), such that 
\be 
&(a - 1) \dot \theta \, 
\theta - \frac{1}{2} (a - 1) [ (\, \dot \theta )^2 
+ ( \, \theta^\prime )^2 ] 
\non \\
& = \,- \, \dot \theta \, \theta^\prime 
- \frac{1}{2} (a - 1) [ \frac{1}{(a-1)^2}(\, \theta^\prime )^2 
+ ( \, \theta^\prime )^2 ]\,, 
\ee 
\no we obtain from (\ref{wza2}), 
\be 
\check S_{WZ}[\theta, A_\mu ] &\rightarrow &
S_{WZ}[\theta, A_\mu ] 
\non\\
&&= \int d^2 x 
\Big\{ -\dot\theta\, \theta^\prime -\frac 12 
\beta(\, \theta^\prime)^2 
+\, e\,\beta\, A_1\, \theta^\prime\,+\,2\,e\,A_0\, \theta^\prime \Big\}\,, 
\ee 
\no in agreement with the minimal WZ action, eq. (\ref{WZA}), 
obtained in the previous section.


\section{Concluding Remarks} 
\renewcommand{\theequation}{4.\arabic{equation}} 
\setcounter{equation}{0} 

We have obtained the extended gauge invariant version of the minimal 
chiral Schwinger model (for $a\not=1$) by using the BFFT method. 
Consequently, this gauge invariance might suggest that this version 
of the model (possibly with a convenient choice of the Jackiw-Rajaraman 
regularization parameter $a$) has some correspondence with a bonafide 
gauge model, such as the $QED_2$, which is naturally gauge invariant, 
or in other words, is not gauge anomalous since in this case the anomaly 
resides in the axial current. 

Indeed, it has been argued 
by Kye {\sl et al.} \cite{Kye} that in the case $a=2$ there is a 
correspondence between the GI formulation of the minimal chiral model 
and the $QED_2$ (see also \cite{Carena}). From our point of view, 
this limit in the standard chiral models as well as in the minimal 
chiral models, does not physically represent the (vector) Schwinger model, 
in contrast to the equivalence advocated in the literature 
\cite{Kye,Carena}. 

In Ref. \cite{BR} the fermionization of the standard WZ Lagrangian has 
been performed for general Jackiw-Rajaraman parameter $a > 1$. 
The fermionized version of the WZ Lagrangian can be written as a 
Thirring model plus a coupling of the gauge field with the axial and 
vector currents. For the special value $a = 2$ the Thirring coupling 
vanishes and the total Lagrangian of the GI version can be related to 
a model exhibiting some resemblance with the $QED_2$. However, this 
mapping only has a formal character 
since it only can be performed in the Lagrangian level. 
From the functional integral approach, this mapping only can be 
performed into the partition function. 
These correspondences cannot be established for the generating 
functionals, which implies that there is no isomorphism between the 
corresponding Hilbert space of states. In this way, the models cannot be 
considered as being equivalent. As stressed in Refs. 
\cite{Carvalhaes,Bel,BRR}, from the operator point of view, the claimed 
equivalence is a consequence of an improper factorization of the 
Hilbert space that implies the choice of a field operator that does 
not belong to the intrinsic field algebra to represent the 
fermionic content of the model. Contrary to the alleged equivalence, 
we also mention two general properties of two-dimensional anomalous 
gauge theories: $i)$ the anomalous models do not exhibit the violation 
of the asymptotic factorization property (cluster decomposition) and thus 
there is no need of a $\theta$-vacuum parametrization 
\cite{Carvalhaes,Bel,BRR}; $ii)$ the models exhibit a peculiar feature 
which allows two isomorphic formulations: the GNI and GI formulations 
\cite{Carvalhaes,Bel}. The suggested equivalence of the chiral model for 
$a=2$ and the vector model can not be established if we consider the 
Hilbert space in which the intrinsic field algebra of the model is 
represented \cite{AAR,Carvalhaes,Bel}. 


\bigskip 
\bigskip 
\no {\bf Acknowledgments.} The authors were partially supported by 
CNPq -- Brazilian research agency. 


\section*{Appendix A} 
\renewcommand{\theequation}{A.\arabic{equation}} 
\setcounter{equation}{0} 

Here, we are going to calculate $H^{(2)}$, Eq. (\ref{Hc2}). 
First of all, we give some details of the calculation of 
$G_1^{(1)}$ and $G_2^{(1)}$, which are necessary for determining $H^{(2)}$. 
From the definition of $G_\gamma^{(n)}$, Eq. (\ref{Gn}), we have 
\be 
G_1^{(1)}&=&\left\{\chi_1,H^{(1)}\right\} \non\\ &=& 
\theta_2(x)\left\{\chi_1,-\chi_2\right\} 
\left(\partial_1\theta_1(x)\right)\ \frac 1{e^2(a-1)^2} 
\left\{\chi_1,\, e^2 \, (a-1) A_0\right\} \non\\ 
&=& - \frac 1{a-1}\left(\partial_1\theta_1(x)\right) - 
e^2(a-1)\theta_2(x). 
\ee 
\no Also, 
\be 
G_2^{(1)}&=&\left\{\chi_2,H^{(1)}\right\} 
\non\\ &=& - \left[\partial_1\theta_1(x)\right] \frac 1{e^2(a-1)} 
\left\{\chi_2,\chi_2\right\}- \theta_2(x)\left\{\chi_2,\chi_2\right\} 
\non\\ && 
- \theta_1(x)\frac 1{e^2(a-1)} 
\left\{\chi_2,e^2\left[(a-1)\partial_1A_1+\Pi^1\right]\right\} 
\non\\ &=& 
- \frac 2{(a-1)^2}\left(\partial_1\theta_1(x)\right)\partial_{1} 
- \theta_1(x)(\partial_{1})^2 \non\\ && - \frac {e^2}{a-1}\theta_1(x) 
- 2e^2\theta_2(x)\partial_{1}. 
\ee 
\no Now, let us calculate $H^{(2)}$ 
\be 
H^{(2)} 
&=& - \frac 12 \int {\rm d}x\ \theta_1(x) \, \epsilon_{12} \, 
\sigma^{21} G_1^{(1)} 
\non\\ && 
- \frac 12 \int {\rm d}x\ \theta_1(x)\, \epsilon_{12}\, \sigma^{22} 
G_2^{(1)} \non\\ 
&& - \frac 12 \int {\rm d}x\ \theta_2(x) 
\, \epsilon_{21} \, \sigma^{11} G_1^{(1)} 
\ee 
\no where $\sigma^{ij}=(\sigma_{ij})^{-1}$, with $\sigma_{ij}$ 
given by Eq. (\ref{sigma}), so that 
\be 
(\sigma_{ij})^{-1} && =\left(\begin{array}{cc}1 & 0 
\\ \frac 1{e^2(a-1)^2}\partial_{x} & -\frac 1{e^2(a-1)} 
\end{array} \right)\delta(x-y). 
\label{sigma-1} 
\ee 
\no Then, we have 
\be H^{(2)}&=&-\frac 12\int{\rm d}x\ \theta_1(x) 
\frac 1{e^2(a-1)^2} \partial_x 
\Big[\frac 1{a-1}(\partial_1\theta_1(x)) 
+(a-1)e^2\theta_2(x)\Big] 
\non\\ 
&+&\int{\rm d}x\ \theta_1(x) \frac 1{e^2(a-1)} 
\Big[\frac 1{(a-1)^{2}}(\partial_1\theta_1(x))\partial_x 
\non\\ &&\qquad\qquad 
+\frac 12\theta_1(x)(\partial_x)^2 +\frac 12\frac 1{(a-1)e^2}\theta_1(x) 
+\, e^2 \, \theta_2(x)\partial_x\Big] \non\\ &+&\frac 12\int{\rm d}x\ 
\theta_2(x) \Big[\frac 1{a-1}(\partial_1\theta_1(x)) 
+(a-1)e^2\theta_2(x)\Big] 
\ee 
\no Finally, after some algebra and making the shift given by 
Eqs. (\ref{shift1}), (\ref{shift2}), one arrives at $H^{(2)}$, 
Eq. (\ref{Hc2}). 



\end{document}